\documentclass{mpe_report}

\usepackage{psfig,graphicx,epsfig}
\usepackage{color}
\usepackage{amsmath,amssymb,epic,eepic,array}
\usepackage{natbib}

\begin{document}

\pagenumbering{arabic}
\setcounter{page}{96}

 \renewcommand{\FirstPageOfPaper }{ 96}\renewcommand{\LastPageOfPaper }{ 99}\newcommand{\degs}{\mbox{$^{\circ}$}}
\newcommand{\rf}{\par\noindent\hangindent 15pt {}}
\newcommand{\rref}{\par\noindent\hangindent 15pt {}}
\newcommand{\ex}[2]{\mbox{#1$\times$10$^{#2}$}}  
\newcommand{\phn}{\phantom{0}}
\newcommand{\tnmp}{$^{\rm p}$}
\newcommand{\tnmn}{$^{\rm n}$}
\newcommand{\tnmo}{$^{\rm o}$}
\newcommand\nodata{~$\cdots$~}%

\authorrunning{Lommen et al.}


\title{Observed Luminosity Difference Between Isolated and Binary MSPs}

\author{Andrea N. Lommen\inst{1} \and Richard A. Kipphorn\inst{1}
\and
David J. Nice\inst{2}
\and
Eric M. Splaver\inst{3}
\and
Ingrid H. Stairs\inst{4}
\and
Donald C. Backer\inst{5}}
\institute{Department of Physics and Astronomy, Franklin and Marshall College\\
501 Harrisburg Pike, Lancaster, Pennsylvania, 17603
\and
Physics Department, Bryn Mawr College \\
Bryn Mawr, PA  19010
\and
Physics Department, Princeton University \\
Princeton, NJ  08544
\and
Department of Physics and Astronomy, University of British Columbia\\
6224 Agricultural Road, Vancouver, BC V6T 1Z1, Canada
\and
Department of Astronomy and Radio Astronomy Laboratory\\
University of California, Berkeley, CA 94720}

\maketitle

\begin{abstract}

We perform a brief census of velocities of isolated versus binary millisecond
pulsars.  We find the velocities of the two populations are indistinguishable.
However, the scale height of the binary population is twice that of the isolated population
and the luminosity functions of the two populations are different. 
We suggest that the scale height difference
may be an artifact of the luminosity difference.  We examine the
magnetic fields of the two populations as a possible source of the luminosity
difference. 

\end{abstract}

\section{Introduction}

We expect MSPs to have lower velocities than the population of regular pulsars, because the kick from
the supernova progenitor had to be small enough to leave the binary intact.  Producing
an isolated MSP requires an even more particular scenario. The binary must
remain intact during and after the supernova, and then after the spin-up phase the companion
must leave the system or be evaporated.
Several authors have debated whether isolated MSP velocities are lower, higher,
or indistinguishable from those of the general population of MSPs.
\cite{McLaughlin04b} suggest we might expect isolated MSPs to have
higher velocities.
They argue that if isolated MSPs are formed by ablation, we would expect them
to form from the tighter binaries which are more susceptible to ablation.
The correlation between tight binaries and higher velocities is suggested by \cite{Tauris96}.
\cite{McLaughlin04b} present the argument for faster velocities for
isolated MSPs as a counterpoint to
their timing proper motion and scintillation measurements
which suggest the opposite, as do the measurements of
\cite{Johnston98} and \cite{Toscano99}.
\cite{Hobbs05}, however,
find the velocities of the populations to be indistinguishable.
In \cite{Lommen06-0030}
we measured
the transverse velocity of PSR J0030+0451 which is unusually small, even compared to the
isolated MSP population.   In this brief article, we have used this as an excuse to reconsider 
the question of the velocity
of isolated MSPs as compared to the binary MSP population.

\section{MSP velocities}
The 29 MSPs in the Galactic Disk with measured
proper motions are shown in
Table \ref{tab:velocities}.
PSR J1730-2304, which has no measured declination proper motion, has
been included in the table for completeness but has not been included in
any of the following calculations.  We corrected each pulsar's velocity to its
LSR as follows.
We used the measured proper motion and distance to calculate a three
dimensional vector representing the (two dimensional) transverse
motion of the pulsar in the reference frame of the Sun.  We then
removed the solar motion and rotated the resulting vector from the LSR
of the sun to the LSR of the pulsar.  Finally, we recovered those two
components of the vector which are perpendicular to the line of sight.
This computation required selecting a value for the unknown
LOS velocity of the pulsar;  we chose a value appropriate
for a star at rest in the pulsar's LSR.

This corrected
velocity is listed in the second to last column of Table 1.
The
average corrected velocity of the isolated MSPs is $86 \pm 19$ km~s$^{-1}$ whereas
the average corrected velocity of all the binary MSPs is $91 \pm 28$ km~s$^{-1}$.
(The uncorrected averages are 79 and 90 km~s$^{-1}$ respectively).
If one
allows the sample to include only those proper motions which have been measured
to better than 2$\sigma$ the average corrected velocities are
$86 \pm 19$\,km~s$^{-1}$ and $99 \pm 33$\, km~s$^{-1}$ respectively.  (Uncorrected
averages are 79 and 99 km~s$^{-1}$.)  In each case
the isolated MSP population is indistinguishable from the binary MSP population.
The 2$\sigma$ cutoff in velocity introduces a selection of higher velocity
pulsars. Thus, the average velocities are higher in that case.

\begin{figure}
\centerline{\psfig{file=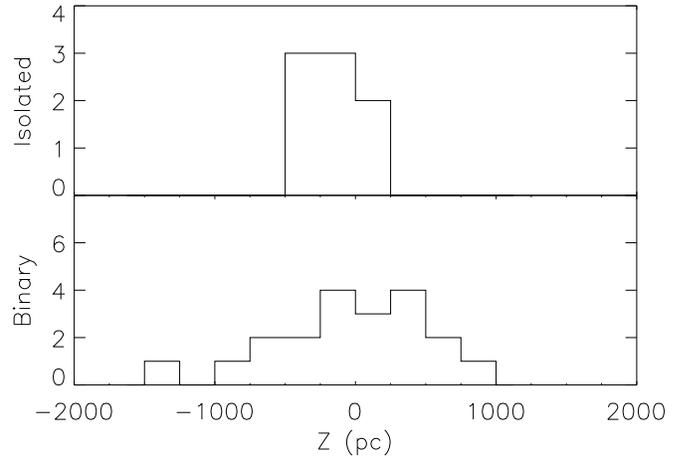,width=8.8cm,clip=} }
\caption{
Histograms of height above the galactic plane for the isolated MSP population (upper) and 
the binary MSP population (lower).
\label{fig:zheight}
}
\end{figure}

An alternative statistic for evaluating the dynamics of pulsar populations
is the distribution of heights above
or below the galactic plane, $z$.  For the pulsars listed, one finds
that the standard deviation from zero for the binary MSP population
is twice that of the isolated MSP population:  $570 \pm 90$\,pc vs
$280 \pm 65$\,pc.
Figure \ref{fig:zheight} shows a histogram of $z$ for each population.  The
isolated MSP population is represented in the upper half of the figure,
the binary MSP population in the lower half.

Figure \ref{fig:zheight}
shows that the known isolated MSPs are closer to the Plane than
are the known binary MSPs.  This could be either a reflection of
differences in the intrinsic spatial distributions of the two types of
MSPs, or a selection effect.  A smaller intrinsic spread in scale
heights for isolated MSPs is only possible if that population also has a
smaller intrinsic velocity distribution, so that the objects do not
travel as far from the Plane as they oscillate in the Galaxy's
potential.  Our determination that the two velocity distributions are in
fact indistinguishable makes this scenario unlikely.  However, with
identical velocity distributions, a difference in intrinsic {\it
luminosity} distributions would cause the less-luminous population to be
detected only to smaller distances and hence only to smaller scale
heights.  In fact, \cite{Bailes97} find that luminosities of
isolated and binary MSPs are different at the 99.5\% confidence level,
with the isolated MSPs being intrinsically dimmer.  We have confirmed
their results with an updated catalog; also, a simple examination of the
median distance of the isolated population (510 pc) compared to
the median distance of the binary population (1155 pc)
suggests that the
isolated MSPs must be less luminous.

\begin{table*}
\caption{
Velocities of Millisecond Pulsars in the Galactic Disk
\label{tab:velocities}
}
\begin{tabular}{r@{}lr@{}llr@{}llrrl}
\hline
\hline
 & {Pulsar} & \multicolumn{3}{c}{$\mu_\alpha$}
& \multicolumn{3}{c}{$\mu_\delta$}
& {Distance} & \multicolumn{1}{c}{$v_t$}
& {Reference}
\\
&  & \multicolumn{3}{c}{(mas~yr$^{-1}$)}
& \multicolumn{3}{c}{(mas~yr$^{-1}$)}
& \multicolumn{1}{c}{(pc)} & \multicolumn{1}{c}{(km~s$^{-1}$)}
& 
\\
\hline
\multicolumn{11}{c}{Isolated MSPs} \\
\hline
J & 0030+0451 & $\mu_\lambda = -5$ &.84 &$\pm$\phn0.09 & $|\mu_\beta| < 10$&& &  310\tnmp &   $<$20 & \cite{Lommen06-0030} \\
J & 0711$-$6830 &    $-$15&.7 &$\pm$\phn0.5 &     15&.3 &$\pm$\phn0.6  &  860\tnmn &  113 & \cite{Toscano99a}\\
J & 1024$-$0719 &      $-$41&&$\pm$\phn2 &     $-$70& &$\pm$\phn3  &  200\tnmo & 70 & \cite{Toscano99a}\\
J & 1730$-$2304 &     20&.5 &$\pm$\phn0.4 &   \multicolumn{2}{c}{\nodata}&&  510\tnmn &  $>$50 & \cite{Toscano99a}\\
J & 1744$-$1134 &    18&.64 &$\pm$\phn0.08 &    $-$10&.3 &$\pm$\phn0.5  &  360\tnmp &  31 & \cite{Toscano99a}\\
  B & 1937+21 &    $-$0&.130 &$\pm$\phn0.008 &    $-$0&.469 &$\pm$\phn0.009 & 3600\tnmn &  87 & \cite{Kaspi94}\\
J & 1944+0907 &     12&.0 &$\pm$\phn0.7 &      $-$18& &$\pm$\phn3  & 1800\tnmn & 173 & \cite{Champion05}\\
J & 2124$-$3358 &      $-$14&&$\pm$\phn1 &      $-$47& &$\pm$\phn1  &  270\tnmn &  48 & \cite{Toscano99a}\\
J & 2322+2057 &      $-$17&&$\pm$\phn2 &      $-$18& &$\pm$\phn3  &  790\tnmn &  79 & \cite{NiceTaylor95}\\
\hline
\multicolumn{11}{c}{Binary MSPs} \\
\hline
J & 0437$-$4715 &  121&.438 &$\pm$\phn0.006 &      $-$71&.438 &$\pm$\phn0.007  &  140\tnmp &  84 & \cite{vanStraten01}\\
J & 0613$-$0200 &      2&.0 &$\pm$\phn0.4 &       $-$7& &$\pm$\phn1  & 1700\tnmn &  60 & \cite{Toscano99a}\\
J & 0751+1807	&$\mu_\lambda = 0$&.35&$\pm$\phn0.03& $\mu_{\beta}=-6$&&$\pm$\phn 2 	& 1150\tnmn	&  22	& \cite{Nice05}\\
J & 1012+5307 &      2&.4 &$\pm$\phn0.2 &    $-$25&.2 &$\pm$\phn0.2  &  840\tnmo &  107 & \cite{Lange01}\\
J & 1045$-$4509 &       $-$5&&$\pm$\phn2 &        6& &$\pm$\phn1  & 1940\tnmn &  119 & \cite{Toscano99a}\\
J & 1455$-$3330 &        5&&$\pm$\phn6 &       24& &$\pm$12  &  530\tnmn &  71 & \cite{Toscano99a}\\
J & 1640+2224 &     1&.66 &$\pm$\phn0.12 &    $-$11&.3 &$\pm$\phn0.2  & 1160\tnmn &  67 & \cite{Loehmer05}\\
J & 1643$-$1224 &        3&&$\pm$\phn1 &       $-$8& &$\pm$\phn5  & 2320\tnmn &  96 & \cite{Toscano99a}\\
J & 1709+2313 &     $-$3&.2 &$\pm$\phn0.7 &     $-$9&.7 &$\pm$\phn0.9  & 1390\tnmn &  57 & \cite{Lewandowski04}\\
J & 1713+0747 &      4&.917 &$\pm$\phn0.004 &     $-$3&.933 &$\pm$\phn0.010 & 1100\tnmp &  30 & \cite{Splaver05}\\
B & 1855+09 &    $-$2&.94 &$\pm$\phn0.04 &    $-$5&.41 &$\pm$\phn0.06 &  910\tnmp &  17 & \cite{Kaspi94}\\
J & 1909$-$3744 &     $-$9&.6 &$\pm$\phn0.2  &    $-$35&.6 &$\pm$\phn0.7  &  820\tnmp & 131 & \cite{Jacoby03}\\
J & 1911$-$1114 &       $-$6&&$\pm$\phn4 &      $-$23& &$\pm$13  & 1220\tnmn & 128 & \cite{Toscano99a}\\
B & 1953+29 &     $-$1&.0 &$\pm$\phn0.3 &     $-$3&.7 &$\pm$\phn0.3  & 4610\tnmn &  128 & \cite{Wolszczan00_4msp}\\
B & 1957+20 &    $-$16&.0 &$\pm$\phn0.5 &    $-$25&.8 &$\pm$\phn0.6  & 2490\tnmn & 325 & \cite{Arzoumanian94}\\
J & 2019+2425 &    $-$9&.41 &$\pm$\phn0.12 &    $-$20&.60 &$\pm$\phn0.15  &  1490\tnmn &  142 & \cite{Nice01}\\
J & 2051$-$0827 &       1& &$\pm$\phn2 &      $-$5& &$\pm$\phn3  & 1040\tnmn &  42 & \cite{Stappers98}\\
J & 2129$-$5721 &        7&&$\pm$\phn2 &       $-$4& &$\pm$\phn3  & 1340\tnmn &  48 & \cite{Toscano99a}\\
J & 2229+2643 &      1& &$\pm$\phn4 &      $-$17& &$\pm$\phn4  & 1440\tnmn & 130 & \cite{Wolszczan00_4msp}\\
J & 2317+1439 &     $-$1&.7 &$\pm$\phn1.5 &      7&.4 &$\pm$\phn3.1  & 820\tnmn &  20 & \cite{Camilo96}\\
\hline
\end{tabular}
\begin{list}{}{}
\item[$^{\rm p}$] Distance from parallax.
\item[$^{\rm n}$] DM distance from NE2001.
\item[$^{\rm o}$] Some other method used to acquire distance.  The text of the cited reference should
be consulted for details.  
\end{list}
\end{table*}
 
\section{Magnetic field}
If the luminosity difference is real then perhaps we could actually observe a magnetic
field difference between the populations.  We did a ``quick and dirty" census of
magnetic fields simply using
the Parkes pulsar catalog\footnote{http://www.atnf.csiro.au/research/pulsar/psrcat/}.
Figure 2 shows a histogram of the magnetic fields of the two populations.  As you can
see, our brief investigation into this matter was inconclusive.  We intend to do
a more complete study in the near future.

\begin{figure}
\centerline{\psfig{file=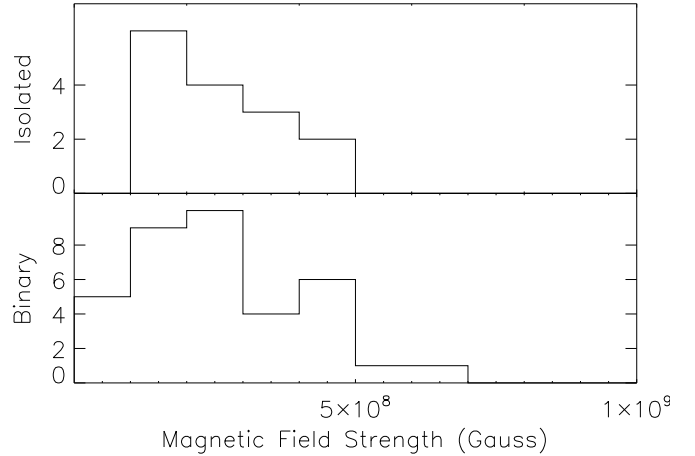,width=8.8cm,clip=} }
\caption{
A histogram of isolated and binary pulsar magnetic field strength for field 
MSPs in 
which a period derivative is measured.
}
\end{figure}

\section{Conclusion}
We conclude that isolated MSPs are less luminous than binary MSPs.  As a starting
point for a search for the cause of the difference we compared the magnetic
fields of the two populations but found they were similar.  Ablation is currently
the favored scenario for creation of isolated MSPs, but it is not known
how ablation could render MSPs less luminous or why MSPs more
likely to ablate their companions would be less luminous to begin with.  
Currently we only know of two ablating MSPs in the disk:
PSR B1957+20 and J2051-0827 \cite{Stappers03, Stappers98}.
The discovery
of more MSPs, and in particular of more ablating MSPs will aid
finding the source of the difference in luminosities.

\begin{acknowledgements}
The Arecibo Observatory is a facility of the National
Astronomy and Ionosphere Center, operated by Cornell University under 
a cooperative agreement with the National Science
Foundation (NSF).   
ANL acknowledges a Research Corporation award in support of this research.  
DJN is supported by NSF grant AST-0206205. 
IHS holds an NSERC UFA and is supported by a Discovery Grant.
DCB acknowledges support from NSF AST-9987278 for ABPP instrumentation and
NSF AST-0206044 for the science program.
\end{acknowledgements}


          \clearpage

\end{document}